\documentclass[
prd,
preprint,
superscriptaddress,
tightenlines,
showpacs
,showkeys
,nofootinbib
,aps
,amsfonts
,amssymb
]{revtex4}

\usepackage{amsmath}
\usepackage{graphicx}
\usepackage{epic}
\usepackage{eepic}
\usepackage{epsfig}
\usepackage{latexsym}
\usepackage{subfigure}
\usepackage{float}
\usepackage[all]{xy}
\usepackage{amsfonts}
\usepackage{amssymb} 
\usepackage{color}
\usepackage{multirow} 
\newcommand{\bce}{\begin{center}}  
\newcommand{\ece}{\end{center}}
\newcommand{\beq}{\begin{equation}}  
\newcommand{\eeq}{\end{equation}}
\newcommand{\beqy}{\begin{eqnarray}}
\newcommand{\eeqy}{\end{eqnarray}}

\def\){\right)} 
\def\({\left(} 
\def\]{\right]} 
\def\[{\left[}

\begin{document}

\title{Neutrality of a magnetized two-flavor quark superconductor}

\author{Tanumoy Mandal}
\email{tanumoy@imsc.res.in}
\affiliation{The Institute of Mathematical Sciences, Chennai, TN 600113, India}
\author{Prashanth Jaikumar}
\email{pjaikuma@csulb.edu}
\affiliation{California State University Long Beach, Long Beach, CA 90840 USA}

\begin{abstract} 

We investigate the effect of electric and color charge neutrality on the two-flavor color superconducting (2SC) phase of cold and dense quark matter in presence of constant external magnetic fields and at moderate baryon densities. Within the framework of the Nambu-Jona-Lasinio (NJL) model, we study the inter-dependent evolution of the quark's BCS gap and constituent mass with increasing density and magnetic field. While confirming previous results derived for the highly magnetized 2SC phase with color neutrality alone, we obtain new results as a consequence of imposing charge neutrality. In the charge neutral gapless 2SC phase (g2SC), a large magnetic field drives the color superconducting phase transition to a crossover, while the chiral phase transition is first order. At larger diquark-to-scalar coupling ratio $G_D/G_S$, where the 2SC phase is preferred, we see hints of the Clogston-Chandrasekhar limit at a very large value of the magnetic field ($B\sim 10^{19}$G), but this limit is strongly affected by Shubnikov de Haas-van Alphen oscillations of the gap, indicating the transition to a domain-like state. 

\end{abstract}
\pacs{26.60.-c, 24.85.+p, 97.60.Jd}
\keywords{quark matter, color superconductivity, neutron stars}

\maketitle 
\section{Introduction}
\label{sec_intro}
Color superconductivity/superfluidity \cite{Bailin:1983bm,Rapp:1997zu,Alford:1998mk,Rajagopal:2000wf,Reuter:2004ze,Schmitt:2004hg,Shovkovy:2004me,Buballa:2005mv} is now a generally accepted conjecture about the state of matter at high baryon densities and low temperatures. Many of its features have been quite comprehensively explored, and are summarized or referenced in a recent review~\cite{Alford:2007xm}. The phenomenon describes Cooper pairing between quarks, with specific pairing patterns predicted as a function of ``control" parameters such as external fields, quark masses or momentum mismatch, the latter arising usually as a consequence of neutrality conditions~\cite{Alford:2000ze,Rajagopal:2000ff,Alford:2002kj,Steiner:2002gx,Ferrer:2005vd,Neumann:2002jm,Shovkovy:2004um, Sharma:2006ig}. For example, at very high density $\mu_q\gg \Lambda_{\rm QCD}$ and for $N_f=3$ massless flavors without any magnetic fields, the preferred pairing pattern is a flavor and color-democratic one termed the color-flavor-locked (CFL) phase~\cite{Alford:1998mk}. As we tune parameters such as quark masses or magnetic fields to values typically found in compact stars, the CFL state would yield to a less symmetric pairing pattern. It is important to study these different manifestations of color superconductivity, since any hope of observing such exotic phases is tied to the physical properties of compact stars such as neutron or possibly quark stars. However, the analysis gets complicated due to the interplay of many effects. Such stars have enormous magnetic fields, of order $10^{12}$G at the surface, but possibly up to $10^{18}$G in the core, which affects the diquark pairing. Also, one may want to consider the partial restoration of chiral symmetry, since core densities are only moderately high compared to that of the CFL phase. Compact objects are, of course, charge neutral (either locally or globally) and color neutral (locally if we consider homogeneous phases) - this imposes additional constraints as we have mentioned. 

\vskip 0.2cm

In this paper, we investigate the effect of electric and color charge neutrality on the two-flavor color superconducting (2SC) phase of cold and dense quark matter in presence of constant external magnetic fields and at moderate baryon densities. Several works have addressed parts of this problem and arrived at important conclusions: we summarize some of them here to orient ourselves along the new directions in this work. In the earliest works, based on expansions of the thermodynamic potential in powers of $M_s/\mu, \Delta/\mu$~\cite{Alford:2002kj,Steiner:2002gx}, it was found that the neutral unmagnetized 2SC phase is energetically disfavored compared to the CFL phase at compact star densities. However, within the NJL-model, it was determined that there are windows of the phase diagram where such a 2SC phase can arise, provided the diquark coupling strength (expressed as a ratio to scalar coupling) $G_D/G_S$ is chosen to be large and quark masses are computed dynamically~\cite{Ruester:2005jc,Buballa:2003qv}. At more natural values of $G_D/G_S$, gapless quasiparticles arise when the mismatch in down and up quark chemical potentials $\delta\mu\equiv \mu_e/2>\Delta$~\cite{Shovkovy:2003uu,Huang:2003xd,Mishra:2004gw}. 
This gapless phase has a chromomagnetic instability~\cite{Huang:2004bg} that may signal a departure from a homogenous superconductor to a heterogeneous one (eg., the LOFF state~\cite{Giannakis:2004pf}). Our analysis, which includes the gapless phase, and omits the possibility of a LOFF phase, must therefore be taken with the caveat that the gapless phase is otherwise stabilized without disruption in the pairing pattern. The introduction of magnetic fields into the quark superconductor has important consequences. Taking the example of CFL phase, one of these is the emergence of a rotated photon that can penetrate superconducting quark matter, so that magnetic flux screening by the Meissner effect is very weak~\cite{Alford:1999pb}. Another is the reduction in symmetry of the CFL phase since only $d$ and $s$ quarks have the same electromagnetic charge~\cite{Ferrer:2005pu,Ferrer:2006vw}. The magnetic field can enhance certain condensates, which changes the superconducting gap from the unmagnetized state~\cite{Fukushima:2007fc}. Strong oscillations in the gap parameter and the magnetization as a function of applied field can make the homogeneous magnetized phase unstable~\cite{Noronha:2007wg,Noronha:2007ps}. These works focused on the CFL phase, which is automatically color and charge neutral. However, neutron star densities span a more moderate range, where two-flavor pairing and neutrality effects play an important role. Electric charge neutrality of the 2SC phase with chiral condensates was studied in~\cite{Huang:2002zd}, but in the absence of magnetic fields or the possibility of gapless phases. In previous work, we studied the effect of a large magnetic field on the chiral and diquark condensates in a regime of moderately dense quark matter~\cite{Mandal:2009uk}. Our focus was on the inter-dependence of the two condensates through non-perturbative quark mass and strong coupling effects, as done for the unmagnetized case in~\cite{Huang:2001yw} but neutrality effects were ignored. In such a case, a mixed broken phase with co-existing chiral and diquark condensate arises~\cite{Huang:2001yw}. We also found that at large $G_D/G_S$, a large magnetic field changes the nature of the phase transitions in this mixed broken phase region. 

\vskip 0.2cm

An extension of our work~\cite{Mandal:2009uk} to include color neutrality was performed in~\cite{Fayazbakhsh:2010gc} and the authors found a change in the order of the phase transition from normal to superconducting quark matter. Other aspects of magnetized color superconducting matter were explored in~\cite{Fayazbakhsh:2010bh} and~\cite{Fayazbakhsh:2012vr}. A 2-flavor model was assumed in these papers, but since the NJL model has no dynamical color fields, the color chemical potential $\mu_8$ has to be introduced by hand, and turns out to be a very small correction to the free energy. Consequently, its effects on competition between phases is also likely to be small, as observed in~\cite{Fayazbakhsh:2010gc}. However, charge neutrality can be a more important factor for the same. In this paper, we therefore turn our attention to a more comprehensive analysis of the effects of neutrality (both color and charge) for magnetized two-flavor quark matter. This is perhaps the phenomenological regime most relevant to neutron stars. We present results for the quark gap, constituent mass and neutralizing charges as a function of density for weak and strong magnetic fields. While we defer the comparison of different phases to future work, this work has important new emergent conclusions, such as the Clogston-Chandrasekhar limit~\cite{Clogs,Chand} for the g2SC phase, the disappearance of the mixed broken phase and the possibility of strongly magnetized domains of neutral but superconducting quark matter in the star.

\vskip 0.2cm

The paper is organized as follows: In section~\ref{lagran}, we state the model NJL Lagrangian and its parameters. In section~\ref{part}, we recast the partition function in terms of interpolating bosonic variables and derive the thermodynamic potential. In section~\ref{gaps}, we obtain the relevant chiral/diquark gaps as well as the neutralizing charges. In section~\ref{numerics}, we discuss our numerical results for the coupled evolution of the condensates as a function of density and magnetic field for a fixed ratio $G_D/G_S$. Our concluding remarks are in section~\ref{conc}.

%%%%%%%%%%%%%%%%%%%%%%%%%%%%%%%%%%%%%%%%%%%%%%%%%%%%%%%%%%%%%%%%%%%%%%%%%%%%%%%

\section{Lagrangian}
\label{lagran}
We employ a NJL type Lagrangian density for two quark flavors ($N_f=2$) applicable 
to scalar and pseudoscalar mesons and scalar diquarks as follows,
\begin{eqnarray}
\mathcal{L} &=& \bar{q}\left[i\gamma^{\mu}\left(\partial_{\mu} - ieQA_{\mu} - igT^8G^8_{\mu}\right) 
+ \hat{\mu}\gamma^{0} - \hat{m}\right]q + G_{S}\left[\left(\bar{q}q\right)^2
+ \left(\bar{q}i\gamma_{5}\vec{\tau}q\right)^2\right]\nonumber\\
&~& + ~G_D\left[\left(\bar{q}i\gamma_{5}\epsilon_{f}\epsilon_{c}q^C\right)
\left(\bar{q}^{C}i\gamma_{5}\epsilon_{f}\epsilon_{c}q\right)\right]
\end{eqnarray}
where the Dirac spinor $q\equiv q_{ia}$ with $i=(1,2)=(u,d)$ and $a=(1,2,3)=(r,g,b)$ 
is a flavor doublet and a color triplet. The charge-conjugate fields of $q$ and 
$\bar{q}$ are defined as $q^C=C\bar{q}^T$ and $\bar{q}^C=-q^TC$ with $C=-i\gamma^0\gamma^2$.
The vector $\vec{\tau}=(\tau^1,\tau^2,\tau^3)$, where the components are the Pauli
matrices in flavor space, whereas $(\epsilon_f)_{ij}$ and $(\epsilon_c)^{ab3}$ 
are antisymmetric matrices in the flavor and color spaces, respectively. We define 
the chemical potential for each flavor and color by $\hat{\mu}=\mu - Q\mu_e + T^3\mu_{3c} + T^8\mu_{8c}$, 
where $Q$ is the generator of $U(1)_{EM}$, $T^3$ and $T^8$ are the two diagonal 
generators of $SU(3)_c$, $\mu$ is the common chemical potential for non-zero 
baryonic density. Since the red and green color of a particular flavor are 
degenerate and the diquark condensates in the blue color direction, we can 
assume $\mu_{3c}=0$. The difference of chemical potentials between the first 
two colored (red and green) quarks and the third colored (blue) quark of a 
same flavor is induced by $\mu_{8c}$, and for a same color, the difference 
of chemical potentials between two flavored quarks ($u$ and $d$) is induced 
by $\mu_e$. The electromagnetic charge matrix for quark defined as $Q=Q_f\otimes \bf 1_c$ 
with $Q_f\equiv\textrm{diag}(2/3,-1/3)$ which coupled to $U(1)$ gauge field 
$A_{\mu}$. Here $e$ is the electromagnetic charge of an electron and $g$ is 
the $SU(3)_c$ coupling constant. The matrix $\hat{m}\equiv \textrm{diag}(m_u,m_d)$ 
is the current quark mass matrix in flavor basis. We take the exact isospin 
symmetry i.e. $m_u=m_d=m_0$. The scalar and diquark couplings are denoted 
as $G_S$ and $G_D$ respectively.
\begin{table}
\begin{center}
\caption{\label{rotcharge}$\tilde{Q}$ charges of quarks in the 2SC phase in units of $\tilde{e}$ in presence of external rotated magnetic field $\tilde{\bf B}$}
\scalebox{1}{
\begin{tabular}{|c|c|c|c|c|c|c|} \hline
\multicolumn{1}{|c|}{Flavor} & \multicolumn{3}{|c|}{up} & \multicolumn{3}{|c|}{down} \\ \cline{1-7} 
\multicolumn{1}{|c|}{Color} & \multicolumn{1}{|c|}{Red} & \multicolumn{1}{c|}{Green} & \multicolumn{1}{c|}{Blue} & \multicolumn{1}{|c|}{Red} & \multicolumn{1}{c|}{Green} & \multicolumn{1}{c|}{Blue}\\ \hline
$\tilde{Q}$-charge  & $+\frac{1}{2}$  & $+\frac{1}{2}$ & 1 & $-\frac{1}{2}$ & $-\frac{1}{2}$ & 0   \\ \cline{1-7} 
\end{tabular}}
\label{table1}
\end{center}
\end{table}

\vskip 0.2cm

We introduce four auxiliary bosonic fields which bosonize the four-fermion 
interactions via a Hubbard-Stratonovich transformation. The bosonic fields are
\begin{eqnarray} 
\sigma = \left(\bar{q}q\right);~~\vec{\pi} = \left(\bar{q}i\gamma_5\vec{\tau}q\right);
~~\Delta = \left(\bar{q}^{C}i\gamma_{5}\epsilon_{f}\epsilon_{c}q\right);
~~\Delta^{*} =\left(\bar{q}i\gamma_{5}\epsilon_{f}\epsilon_{c}q^C\right)
\end{eqnarray}
where $\sigma$ and $\vec{\pi}$ are the mesons and $\Delta$ and $\Delta^{*}$ are 
the diquarks. The bosonized Lagrangian density becomes
\begin{eqnarray}
\mathcal{L} &=& \bar{q}\left[i\gamma^{\mu}\left(\partial_{\mu} - ieQA_{\mu} - igT^8G^8_{\mu}\right) + \hat{\mu}\gamma^{0}\right]q 
- \bar{q}\left(m + i\gamma_5\vec{\pi}\cdot\vec{\tau}\right)q \nonumber\\
&~& - ~ \frac{1}{2}\Delta^{*}\left(\bar{q}^{C}i\gamma_{5}\epsilon_{f}\epsilon_{c}q\right) - \frac{1}{2}\Delta\left(\bar{q}i\gamma_{5}\epsilon_{f}\epsilon_{c}q^C\right)
- \frac{\sigma^2 +
\vec{\pi}^2}{4G_s} - \frac{\Delta^{*}\Delta}{4G_D}
\end{eqnarray}
where $m = m_0 + \sigma$. We do not include the possibility of pion condensation 
for simplicity \cite{Andersen:2007qv}. So we set $\vec{\pi} = 0$ in our analysis. 
In general, one must allow for the flavor dependence of the chiral condensate when isopsin symmetry is broken (in our case by non-zero $\delta\mu$ and magnetic field only, since we assume degenerate light quark masses). This is true even in the absence of color superconductivity (eg.~\cite{Toublan:2003tt}). In our case, with color superconductivity, it is the rotated charge that matters, not the original U(1) charge. The coupling of different quark flavors to the magnetic field is already included consistently in their spectrum (Landau levels) through appropriate rotated charges. The flavor dependence arising from the effect of non-zero $\delta\mu$ has not been studied here, nor in any of the works similar to ours, and would be an interesting feature to explore, but is a separate issue from the effect of the magnetic field which is our focus here. Chiral symmetry breaking and color superconductivity in the 2SC phase is manifest 
by non-vanishing vacuum expectation values (VEV) for $\sigma$ and $\Delta$. In 
this paper we will study the effect of an external constant magnetic field $B$ on
these condensates.

\vskip 0.2cm

Since the diquark condensates of $u$ and $d$ quarks carry a net electromagnetic 
charge, there is a Meissner effect for ordinary magnetism, while a linear 
combination of photon and gluon leads to a rotated massless $U(1)$ field which 
we identify as photon of our theory. We can write the Lagrangian in terms of 
rotated quantities using the following identity,
\begin{eqnarray}
\label{rotquan}
eQA_{\mu} + gT^8G^8_{\mu} = \tilde{e}\tilde{Q}\tilde{A}_{\mu} + \tilde{g}\tilde{T}^8\tilde{G}^8_{\mu}
\end{eqnarray}
In the R.H.S. of the Eq.~(\ref{rotquan}) all quantities are rotated. In 
$flavor\otimes color$ space in units of the rotated charge of an electron 
$\tilde{e}={\sqrt{3}ge}/{\sqrt{3g^2+e^2}}$ the rotated charge matrix is
\begin{eqnarray}
\tilde{Q} = Q_f\otimes{\bf 1}_c - {\bf 1}_f\otimes \frac{T^8_c}{2\sqrt{3}}
\end{eqnarray}
The generator $T^3$ plays no role here because the degeneracy of color $1$ and $2$ 
ensures that there is no long range gluon $3$-field. For 2SC phase the 
rotated $\tilde{Q}$ charges of different quarks are presented in Table~\ref{rotcharge}. 
We take constant rotated background $U(1)$ magnetic field ${\bf B} = B\hat{z}$
along $+z$ axis. After integrating out the dynamical part of $A_{\mu}$, 
the potential becomes $A_{\mu}=(0,0,Bx,0)$ in the Landau gauge. The gapped 
2SC phase is $\tilde{Q}$-neutral, while overall charge neutrality of the 
matter requires a neutralizing background of strange quarks and/or electrons. 
In this article, we take the strange quark mass very large so that they do not 
play any dynamical role.

%%%%%%%%%%%%%%%%%%%%%%%%%%%%%%%%%%%%%%%%%%%%%%%%%%%%%%%%%%%%%%%%%%%%%%%%%%%%%%%%%%%

\section{Partition function}
\label{part}
We write the partition function in the presence of a uniform magnetic field 
in the mean field approximation using the standard finite temperature path 
integral formalism over the quark fields
\begin{eqnarray}
\mathcal{Z} = N \int [d\bar{q}][dq] \textrm{exp} \left\{\int_{0}^{\beta}d\tau\int
d^{3}\vec{x}\left(\tilde{\mathcal{L}} - \frac{B^2}{2}\right)\right\}
\end{eqnarray}
where $N$ is the normalization constant, $\tilde{\mathcal{L}}$ is the rotated
Lagrangian density and $\beta=1/T$ is the inverse of the 
temperature $T$.
We add the kinetic energy contribution of the rotated $U(1)$ external gauge field
$-\frac{1}{4}(F_{\mu\nu})^2 = -\frac{B^2}{2}$ in the partition function. The blue
$u$ and $d$ quarks do not participate in the diquark condensate. Whereas, the red $u$
paired with green $d$ and green $u$ paired with red $d$ to form $\tilde{Q}$-charge neutral
diquark condensates. Thus, we can write the full partition function as a product of
three parts, 
\begin{eqnarray}
\mathcal{Z} = \mathcal{Z}_c\mathcal{Z}_{u_b d_b}\mathcal{Z}_{u_r d_g, u_g d_r}
\end{eqnarray}
where $\mathcal{Z}_{u_b d_b}$ part is for the unpairing quarks and the 
$\mathcal{Z}_{u_r d_g, u_g d_r}$ part is for the quarks participating in pairing.
The explicit form of  $\mathcal{Z}_{u_b d_b}$ and $\mathcal{Z}_{u_r d_g, u_g d_r}$
we present in the next section.
The bosonized part $\mathcal{Z}_c$ serves as a constant multiplicative factor and 
reads as
\begin{eqnarray}
\mathcal{Z}_c = N{\rm exp}\left\{ - \int_0^{\beta}d\tau\int d^3x\left[\frac{\sigma^2}{4G_s}
+ \frac{\Delta^{2}}{4G_D} + \frac{B^{2}}{2}\right]\right\}
\end{eqnarray}

The Lagrangian density $\tilde{\mathcal{L}}$ contains a diagonal $6\otimes 6$
chemical potential matrix $\hat{\mu}$ in $flavor\otimes color$ space.
When we incorporate the electric and color charge neutrality conditions, we 
have to assign different chemical potential for each flavor and color. Then 
the diagonal chemical potential matrix $\hat{\mu}$ can be expressed as
\begin{eqnarray}
\hat{\mu} = \textrm{diag}(\mu_{u_r},\mu_{u_g},\mu_{u_b},\mu_{d_r},\mu_{d_g},\mu_{d_b})
\end{eqnarray}
where $\mu_{i_a}$ is the chemical potential for flavor $i$ and color $a$. The
explicit expressions of all $\mu_{i_a}$s are as follows
\begin{eqnarray}
\mu_{u_r}=\mu_{u_g}=\mu - \frac{2}{3}\mu_e + \frac{1}{3}\mu_{8c};~~
\mu_{u_b}=\mu - \frac{2}{3}\mu_e - \frac{2}{3}\mu_{8c}\nonumber\\
\mu_{d_r}=\mu_{d_g}=\mu + \frac{1}{3}\mu_e + \frac{1}{3}\mu_{8c};~~
\mu_{d_b}=\mu + \frac{1}{3}\mu_e - \frac{2}{3}\mu_{8c}
\end{eqnarray}

For the calculational simplicity, we define the mean chemical potential $\bar{\mu}$
and the difference of the chemical potential $\delta\mu$ as 
\begin{eqnarray}
\bar{\mu} &=&\frac{1}{2}(\mu_{u_r}+\mu_{d_g})=\frac{1}{2}(\mu_{u_g}+\mu_{d_r})
=\mu - \frac{1}{6}\mu_e + \frac{1}{3}\mu_{8c}\\
\delta\mu &=&\frac{1}{2}(\mu_{d_g}-\mu_{u_r})=\frac{1}{2}(\mu_{d_r}+\mu_{u_g})
=\frac{1}{2}\mu_e
\end{eqnarray}

%%%%%%%%%%%%%%%%%%%%%%%%%%%%%%%%%%%%%%%%%%%%%%%%%%%%%%%%%%%%%%%%%%%%%%%%%%%%%%%%%%

In order to evaluate the thermodynamic potential $\Omega = - T\ln Z/V$, we introduce 
Nambu-Gorkov bispinors for each color and flavor of quark and express $\mathcal{Z}_{u_b d_b}$
and $\mathcal{Z}_{u_r d_g, u_g d_r}$ in the following way
\begin{eqnarray}
\ln \mathcal{Z}_{u_b d_b} = \frac{1}{2}\ln\left\{\textrm{Det}(\beta\mathcal{G}_0^{-1})\right\};~~
\ln \mathcal{Z}_{u_r d_g, u_g d_r} = \frac{1}{2}\ln\left\{\textrm{Det}(\beta\mathcal{G}_{\Delta}^{-1})\right\}
\end{eqnarray}

The determinant operation is carried out over the Nambu-Gorkov, color, flavor and 
momentum-frequency space and $\mathcal{G}_0^{-1}$ and $\mathcal{G}_{\Delta}^{-1}$ have
the following form
\begin{eqnarray}
\mathcal{G}_0^{-1} = 
\begin{pmatrix}
& [G_{0,\tilde{Q}}^+]^{-1} & 0 \\
& 0 & [G_{0,-\tilde{Q}}^-]^{-1}
\end{pmatrix};~~~
\mathcal{G}_{\Delta}^{-1} = 
\begin{pmatrix}
& [G_{0,\tilde{Q}}^+]^{-1} & \Delta^{-}\\
&\Delta^{+} & [G_{0,-\tilde{Q}}^-]^{-1}
\end{pmatrix}
\end{eqnarray}
where we have used the notation $\Delta^-(\Delta^+)=-i\gamma_5\epsilon_{f}\epsilon_{c}\Delta(\Delta^*)$.
The Green's functions $[G_{0,\tilde{Q}}^{\pm}]^{-1}$ read as follows
\begin{eqnarray}
\left[G_{0,\tilde{Q}}^{\pm}\right]^{-1} \equiv \left[\gamma^{\mu}\left(i\partial_{\mu} + \tilde{e}\tilde{Q}\tilde{A}_{\mu}
\pm\hat{\mu}\gamma^0\right) - m \right]
\end{eqnarray}

The determinant computation is simplified by re-expressing the $\tilde{Q}$-charges
in terms of charge projectors in the color-flavor basis, following techniques 
applied for the CFL phase~\cite{Noronha:2007wg}. With the color-flavor structure 
unraveled, we can simplify the determinant computation further by introducing energy 
projectors~\cite{Huang:2001yw} and moving from position to momentum space using 
Fourier transformation, whereby we find
\begin{eqnarray}
\label{zchiral}
\ln\mathcal{Z}_{u_b d_b} &=& \sum_{p_0, {\bf p}}
\Big[\ln\left\{\beta^2\left(p_0^2 - {E^{+^2}_{u_b}}\right)\right\}
+ \ln\left\{\beta^2\left(p_0^2 - {E^{-^2}_{u_b}}\right)\right\}\nonumber\\
&~& + \ln\left\{\beta^2\left(p_0^2 - {E^{+^2}_{d_b}}\right)\right\}
+ \ln\left\{\beta^2\left(p_0^2 - {E^{-^2}_{d_b}}\right)\right\}\Big]
\end{eqnarray}
\begin{eqnarray}
\label{zdiquark}
\ln\mathcal{Z}_{u_r d_g, u_g d_r} 
&=& 2 \sum_{p_0, {\bf p}} \Big[\ln\left\{\beta^2\left(p_0^2 - \bar{E}^{+^{2}}_{\Delta^{+}}\right)\right\}
+ \ln\left\{\beta^2\left(p_0^2 - {\bar{E}^{+^{2}}_{\Delta^{-}}}\right)\right\}\nonumber\\
&~& + \ln\left\{\beta^2\left(p_0^2 - {\bar{E}^{-^{2}}_{\Delta^{+}}}\right)\right\}
+ \ln\left\{\beta^2\left(p_0^2 - {\bar{E}^{-^{2}}_{\Delta^{-}}}\right)\right\}\Big]
\end{eqnarray}
%The definitions of various terms in the Eq.~(\ref{zchiral}) and Eq.~(\ref{zdiquark}) %are as follows.
We define $E_a = \sqrt{{\bf p}_{\perp,a}^2 + p_z^2 + m^2}$, if $a=0$ then ${\bf p}_{\perp,0}^2 = p_x^2 + p_y^2$ 
else ${\bf p}_{\perp, a}^2=2|a|\tilde{e}Bn$ for $a=\pm 1/2,1$. Here $a$ denotes
the possible $\tilde{Q}$ charges of the quarks and $n$ labels the Landau levels
originated in presence of magnetic field. We see from the definition of
$E_a$ that $E_{\frac{1}{2}}=E_{-\frac{1}{2}}$ and the various other terms in the Eq.~(\ref{zchiral}) and Eq.~(\ref{zdiquark}) are defined as
\begin{eqnarray}
{E^{\pm}_{u_b}} = E_1 \pm \mu_{u_b};~~{E^{\pm}_{d_b}} = E_0 \pm \mu_{d_b}
\end{eqnarray}
\begin{eqnarray}
\bar{E}^{\pm}_{\frac{1}{2}} = E_{\frac{1}{2}}\pm \bar{\mu};~~
\bar{E}^{\pm}_{\Delta} = \sqrt{\bar{E}_{\frac{1}{2}}^{\pm^{2}} + \Delta^{2}};~~
\bar{E}^{\pm}_{\Delta^{\pm}} = \bar{E}^{\pm}_{\Delta} \pm \delta\mu
\end{eqnarray}
The sum over $p_0(=i\omega_k)$ denotes the discrete sum over the Matsubara
frequencies. Using the following identity we can perform the discrete summation 
over $p_0$.
\begin{eqnarray}
\sum_{p_0}\ln\left[\beta^2\left(p_0^2 - E^2\right)\right] = 
\beta\left[E + 2T\ln\left(1 + e^{-\beta E}\right)\right]
\equiv \beta \phi\left(E\right)
\end{eqnarray}
where we have dropped an infinite constant on the left hand side of the equation that is temperature independent, coming from the Matsubara sum (eg~\cite{Huang:2002zd}). We drop this infinite constant as we are only interested in derivatives of the free energy for the gap equations (or differences in free energy from the normal phase, where this constant will cancel). Then, going over to the 3-momentum continuum using the following replacement
\begin{eqnarray}
\sum_{\bf p}\rightarrow V\int\frac{d^3{\bf p}}{{(2\pi)}^3}
\end{eqnarray}
where $V$ is the thermal volume of the system, we can express the zero-field thermodynamic potential as
\begin{eqnarray}
\Omega_{B=0} &=& \frac{m^{2}}{4G_{S}}+\frac{\Delta^{2}}{4G_{D}} + \Omega_{e} - \int_0^{\infty}\frac{d^3{\bf p}}{{(2\pi)}^3}
\Big[ \phi\left(E_{u_b}^{+}\right) + \phi\left(E_{u_b}^{-}\right)
+  \phi\left(E_{d_b}^{+}\right)  \nonumber\\
&~& + \phi\left(E_{d_b}^{-}\right) + 2\left\{ \phi\left( \bar{E}^{+}_{\Delta^{+}} \right)
+ \phi\left( \bar{E}^{+}_{\Delta^{-}} \right)
+ \phi\left( \bar{E}^{-}_{\Delta^{+}} \right)
+ \phi\left( \bar{E}^{-}_{\Delta^{-}} \right) \right\}\Big]
\end{eqnarray}
where $\Omega_e$ is the contribution from non-interacting electron gas. In the limit of 
electron zero mass and small $T$, $\Omega_e$ takes the form
\begin{eqnarray}
\Omega_{e} = - \left(\frac{\mu^{4}_e}{12\pi^2} + \frac{\mu^{2}_e T^2}{6} 
+ \frac{7\pi^2 T^4}{180}\right)
\end{eqnarray}

In presence of a quantizing magnetic field, discrete Landau levels suggest the following replacement
\begin{eqnarray}
\label{replace}
\int_0^{\infty}\frac{d^3{\bf p}}{{(2\pi)}^3} \rightarrow \frac{|a|\tilde{e}B}{8\pi^{2}}\displaystyle\sum_{n=0}^{\infty}\alpha_{n}\int_{-\infty}^
{\infty}dp_{z}~~{\rm where}~~\alpha_{n}=2-\delta_{n0}\,,
\end{eqnarray}
where $\alpha_n$ is the degeneracy factor of the Landau levels (all levels
are doubly degenerate except the zeroth Landau level). The thermodynamic potential 
in presence of a magnetic field is given by
\begin{eqnarray}
\label{omegaB}
\Omega_{B} = \frac{m^{2}}{4G_{S}}+\frac{\Delta^{2}}{4G_{D}} +\frac{B^2}{2}+ \Omega_{e}
+ \sum_{a} \Omega_{a}~~\textrm{where}~~a\in 0,1,\frac{1}{2}
\end{eqnarray}
The $\Omega_a$ is the contribution to the $\Omega_{B}$ from the quarks of 
$\tilde{Q}$-charge $a$ and they are
\begin{eqnarray}
\Omega_{0} = - \int_0^{\infty}\frac{d^3{\bf p}}{{(2\pi)}^3}
\Big[ \phi\left(E_{d_b}^{+}\right) + \phi\left(E_{d_b}^{-}\right)\Big]
\end{eqnarray}
\begin{eqnarray}
\Omega_{1} = - \frac{\tilde{e}B}{8\pi^{2}}\displaystyle\sum_{n=0}^{\infty}\alpha_{n}
\int_{-\infty}^{\infty}dp_{z}
\Big[ \phi\left(E_{u_b}^{+}\right) + \phi\left(E_{u_b}^{-}\right)\Big]
\end{eqnarray}
\begin{eqnarray}
\Omega_{\frac{1}{2}} = - \frac{\tilde{e}B}{8\pi^{2}}\displaystyle\sum_{n=0}^{\infty}\alpha_{n}
\int_{-\infty}^{\infty}dp_{z}
\Big[\phi\left( \bar{E}^{+}_{\Delta^{+}} \right)
+ \phi\left( \bar{E}^{+}_{\Delta^{-}} \right)
+ \phi\left( \bar{E}^{-}_{\Delta^{+}} \right)
+ \phi\left( \bar{E}^{-}_{\Delta^{-}} \right)\Big]
\end{eqnarray}

%%%%%%%%%%%%%%%%%%%%%%%%%%%%%%%%%%%%%%%%%%%%%%%%%%%%%%%%%%%%%%%%%%%%%%%%%%%%%%%%%%

\section{Gap equations and neutrality conditions}
\label{gaps}
Minimizing the thermodynamic potential in Eq.~(\ref{omegaB}) w.r.t. $m$ and $\Delta$,
we can derive chiral and diquark gap equations respectively. We can also get the 
color and electric charge neutrality conditions minimizing the thermodynamic potential
w.r.t. $\mu_{8c}$ and $\mu_{e}$ respectively.

\subsection{Chiral gap equation}

The gap equation for chiral condensate can be obtained from the following equation
\begin{eqnarray}
\frac{\partial\Omega_{B}(\sigma,\Delta,\mu_{8c},\mu_{e};\mu,\tilde{e}B,T)}{\partial\sigma} = 0
\end{eqnarray}
\begin{eqnarray}
\label{chigap}
\frac{\partial\Omega_{B}}{\partial\sigma} &=&
\frac{\sigma}{4G_S} - m\bigg[\int_0^{\infty}\frac{d^3{\bf p}}{{(2\pi)}^3}\frac{1}{E_0}\lbrace 1 
- f(E_{d_b}^{+}) - f(E_{d_b}^{-})\rbrace\nonumber\\
&~&-~ \frac{\tilde{e}B}{8\pi^{2}}\displaystyle\sum_{n=0}^{\infty}\alpha_{n}
\int_{-\infty}^{\infty}dp_{z}\frac{1}{E_1}\lbrace 1 
- f(E_{u_b}^{+}) - f(E_{u_b}^{-})\rbrace\nonumber\\
&~&-~ \frac{\tilde{e}B}{16\pi^{2}}\displaystyle\sum_{n=0}^{\infty}\alpha_{n}
\int_{-\infty}^{\infty}dp_{z}\frac{1}{E_{\frac{1}{2}}}\Big[2\frac{\bar{E}^{-}}{\bar{E}^{-}_{\Delta}}\lbrace 1 
- f(\bar{E}^{-}_{\Delta^{+}}) - f(\bar{E}^{-}_{\Delta^{-}})\rbrace\nonumber\\
&~&+~ 2\frac{\bar{E}^{+}}{\bar{E}^{+}_{\Delta}}\lbrace 1 
- f(\bar{E}^{+}_{\Delta^{+}}) - f(\bar{E}^{+}_{\Delta^{-}})\rbrace\Big]\bigg] = 0
\end{eqnarray}
where $f(x) = \left[1 + \exp\left(\beta x\right)\right]^{-1}$ is the Fermi-Dirac distribution function. 

\subsection{Diquark gap equation}

The gap equation for diquark condensate can be obtained from the following equation
\begin{eqnarray}
\frac{\partial\Omega_{B}(\sigma,\Delta,\mu_{8c},\mu_{e};\mu,\tilde{e}B,T)}{\partial\Delta} = 0
\end{eqnarray}
\begin{eqnarray}
\label{diqgap}
\frac{\partial\Omega_{B}}{\partial\Delta} &=&
\Delta\bigg[\frac{1}{4G_D} - 
\frac{\tilde{e}B}{16\pi^{2}}\displaystyle\sum_{n=0}^{\infty}\alpha_{n}
\int_{-\infty}^{\infty}dp_{z}
\Big[2\frac{1}{\bar{E}^{-}_{\Delta}}\lbrace 1 
- f(\bar{E}^{-}_{\Delta^{+}}) - f(\bar{E}^{-}_{\Delta^{-}})\rbrace\nonumber\\
&~&+~ 2\frac{1}{\bar{E}^{+}_{\Delta}}\lbrace 1 
- f(\bar{E}^{+}_{\Delta^{+}}) - f(\bar{E}^{+}_{\Delta^{-}})\rbrace\Big]\bigg] = 0
\end{eqnarray}

\subsection{Color charge neutrality}

The total system should be color neutral. We can impose the color charge neutrality condition by 
choosing $\mu_{8c}$ in such a way that the system has net color charge $n_{8c}$ zero. It follows that,
\begin{eqnarray}
n_{8c} = -\frac{\partial\Omega_{B}(\sigma,\Delta,\mu_{8c},\mu_{e};\mu,\tilde{e}B,T)}{\partial\mu_{8c}} = 0
\end{eqnarray}
\begin{eqnarray}
\label{colneu}
\frac{\partial\Omega_{B}}{\partial\mu_{8c}} &=&
\int_0^{\infty}\frac{d^3{\bf p}}{{(2\pi)}^3}\lbrace f(E_{d_b}^{+}) - f(E_{d_b}^{-})\rbrace
+\frac{\tilde{e}B}{8\pi^{2}}\displaystyle\sum_{n=0}^{\infty}\alpha_{n}
\int_{-\infty}^{\infty}dp_{z}\lbrace f(E_{u_b}^{+}) - f(E_{u_b}^{-})\rbrace\nonumber\\
&~&-~ \frac{\tilde{e}B}{16\pi^{2}}\displaystyle\sum_{n=0}^{\infty}\alpha_{n}
\int_{-\infty}^{\infty}dp_{z}\Big[\frac{\bar{E}^{-}}{\bar{E}^{-}_{\Delta}}\lbrace 1 
- f(\bar{E}^{-}_{\Delta^{+}}) - f(\bar{E}^{-}_{\Delta^{-}})\rbrace\nonumber\\
&~&-~ \frac{\bar{E}^{+}}{\bar{E}^{+}_{\Delta}}\lbrace 1 
- f(\bar{E}^{+}_{\Delta^{+}}) - f(\bar{E}^{+}_{\Delta^{-}})\rbrace\Big] = 0
\end{eqnarray}

\subsection{Electric charge neutrality}

The total system should be electrically neutral. We can impose the electric charge neutrality condition by 
choosing $\mu_{e}$ in such a way that the system has net electric charge $n_{e}$ zero. It follows that,
\begin{eqnarray}
n_{e} = -\frac{\partial\Omega_{B}(\sigma,\Delta,\mu_{8c},\mu_{e};\mu,\tilde{e}B,T)}{\partial\mu_{e}} = 0
\end{eqnarray}
\begin{eqnarray}
\label{eleneu}
\frac{\partial\Omega_{B}}{\partial\mu_{e}} &=&
-\int_0^{\infty}\frac{d^3{\bf p}}{{(2\pi)}^3}\big[2\lbrace f(E_{d_b}^{+}) - f(E_{d_b}^{-})\rbrace\big]
+\frac{\tilde{e}B}{8\pi^{2}}\displaystyle\sum_{n=0}^{\infty}\alpha_{n}
\int_{-\infty}^{\infty}dp_{z}\big[4\lbrace f(E_{u_b}^{+}) - f(E_{u_b}^{-})\rbrace\big]\nonumber\\
&~&+~ \frac{\tilde{e}B}{16\pi^{2}}\displaystyle\sum_{n=0}^{\infty}\alpha_{n}
\int_{-\infty}^{\infty}dp_{z}\Big[2\frac{\bar{E}^{-}}{\bar{E}^{-}_{\Delta}}\lbrace 1 
- f(\bar{E}^{-}_{\Delta^{+}}) - f(\bar{E}^{-}_{\Delta^{-}})\rbrace - 2\frac{\bar{E}^{+}}{\bar{E}^{+}_{\Delta}}
\lbrace 1 - f(\bar{E}^{+}_{\Delta^{+}})\nonumber\\
&~&-~ f(\bar{E}^{+}_{\Delta^{-}})\rbrace
- 6\lbrace f(\bar{E}^{-}_{\Delta^{+}}) + f(\bar{E}^{+}_{\Delta^{+}})
- f(\bar{E}^{+}_{\Delta^{-}}) - f(\bar{E}^{-}_{\Delta^{-}})\rbrace\Big] + \frac{\mu_e^{3}}{\pi^{2}} = 0
\end{eqnarray}

We derive two gap equations and two charge neutrality conditions above for
finite $T$, but our all numerical results in the following section is in the 
$T\rightarrow 0$ limit. We investigate the finite $T$ results elsewhere.

%%%%%%%%%%%%%%%%%%%%%%%%%%%%%%%%%%%%%%%%%%%%%%%%%%%%%%%%%%%%%%%%%%%%%%%%%%%%%%%%%%

\section{Numerical analysis}
\label{numerics}
In order to assess the effects of electric and color charge neutrality on quark 
matter in presence of an external magnetic field, we solve the two gap equations 
with electric and color charge neutrality conditions imposed. Since this set 
of four equations all involve integrals that diverge in the ultra-violet,
we regularize in order to obtain physically meaningful results. It is common 
practice to use a sharp cutoff \footnote{An example is the step function 
$\theta\left(\Lambda - |{\bf p}|\right)$ in momentum space where $\Lambda$ 
is the cutoff momentum. We checked our main numerical results for different cutoff schemes including various smooth cutoff parameterizations and found very little sensitivity. There is also the proper time regularization method, which was employed in obtaining the finite temperature results of hot magnetized 2SC quark matter~\cite{Fayazbakhsh:2010bh}. Their zero temperature results are very similar to a previous work~\cite{Fayazbakhsh:2010gc} that investigated cold magnetized 2SC matter using a smooth cutoff scheme. Thus, we prefer to use a simple smooth cutoff for our $T=0$ case.} to regulate these divergences, but this can introduce abnormal behavior 
in many thermodynamical quantities, especially when a system with a discrete 
energy spectrum is considered. In this numerical study, we therefore use a 
smooth cutoff function while performing the numerical integrations in momentum 
space. There is a variety of smooth regulators used in the literature
\cite{Alford:1998mk, Berges:1998rc, Noronha:2007wg}, and all are expected to 
yield qualitatively similar results. We use a Fermi-Dirac type of smooth regulator
\cite{Fukushima:2007fc} 
\begin{equation}
\label{cutoff}
f_c(p_a) = \frac{1}{2}\left[ 1 - \tanh\left\lbrace\left( \frac{p_a - \Lambda}{\alpha}\right)\right\rbrace\right]
\end{equation}
where $p_a = \sqrt{{\bf p}_{\perp,a}^2 + p_z^2}$, if $a=0$ then 
${\bf p}_{\perp,0}^2 = p_x^2 + p_y^2$ 
else ${\bf p}_{\perp, a}^2=2|a|\tilde{e}Bn$ for $a = 1,\pm 1/2$.
In Eq. (\ref{cutoff}) $\Lambda$ is the momentum cutoff and $\alpha$ is a 
free parameter that measures the sharpness of the 
cutoff scheme. Larger values of $\alpha$ imply that the cutoff is gentler.
We choose $\alpha = 0.01\Lambda$ throughout our numerical analysis. We fix 
a momentum cutoff $\Lambda$ and NJL model parameters in the chiral limit
$m_0=0$ as given in \cite{Huang:2002zd},
%We set  and chiral coupling strength $G_S$  as follows,
\begin{eqnarray}
\label{njlparam}
\Lambda = 0.6533~\textrm{GeV},~~G_S = 5.0163~{\textrm{GeV}}^{-2}~~\textrm{and}~~G_D=\rho G_S
\end{eqnarray}
where $\rho$ is a free parameter. Fierz transforming one gluon exchange
implies $\rho=0.75$ for number of colors $N_c=3$ and fitting the vacuum
baryon mass gives $\rho=2.26/3$ \cite{Ebert:1991pz}. Since the underlying
interaction at moderate density is bound to be more complicated, we
therefore choose to vary the strength of the diquark channel to 
investigate the effect of $\rho$ on the magnetic color superconducting phase. 
In \cite{Huang:2003xd} where no magnetic field is considered,
it is found that the ground state of charge neutral two flavor quark
matter is very sensitive to $G_D$. In particular, one finds the 2SC phase
(i.e. $\Delta > \delta\mu$) for $\rho\gtrsim 0.8$, the g2SC phase 
(i.e. $\Delta < \delta\mu$) for $0.7 \lesssim\rho\lesssim 0.8$ and the normal
quark matter (i.e. $\Delta = 0$) for $\rho \lesssim 0.7$. We take the bare 
quark mass $m_0=0$ throughout our analysis. One can consider realistic 
nonzero $m_0$ and get a set of model parameters by fitting three vacuum 
quantities, namely the pion mass $m_{\pi}$, the pion decay constant $f_{\pi}$ 
and the constituent quark mass $m(\mu = 0)$. For nonzero magnetic field, 
the model parameters will differ only slightly from zero field values. 
For our analysis, we keep model parameters fixed at zero field values. 
We note that for the choice of a very sharp cutoff function, Eq. (\ref{replace}) becomes
\begin{eqnarray}
\int_{0}^{\infty}\frac{d^3{\bf p}}{{(2\pi)}^3}f_c(p)\sim
\int_{0}^{\Lambda}\frac{d^3{\bf p}}{{(2\pi)}^3}\rightarrow \frac{|a|\tilde{e}B}{8\pi^{2}}\displaystyle\sum_{n=0}^{n_{max}}\alpha_{n}\int_{-\Lambda^{\prime}}^
{\Lambda^{\prime}}dp_{z}~~{\rm where}~~n_{max} = {\rm Int}\left[\frac{\Lambda^2}{2|a|\tilde{e}B}\right]
\end{eqnarray}
and $\Lambda^{\prime} = \sqrt{\Lambda^{2}-2|a|\tilde{e}Bn}$. Using the fact that
$p_{z}^{2}\geq 0$, we can find the maximum number of completely occupied
Landau levels $n_{max}$. In the limit of weak magnetic field $n_{max}$ becomes 
very large and the summation over discrete landau levels becomes quasi-continuous. The zero magnetic field results remain an excellent approximation in the weak 
magnetic field limit, as we see below. 

\vskip 0.2cm

We reiterate that our main aim in this work is to investigate the effect of 
magnetic fields on the chiral ($m$) and diquark ($\Delta$) condensates for 
different pairing strengths ($\rho$) and with the electric and the color 
charge neutrality conditions imposed. Whether the quark matter is in g2SC 
phase or 2SC phase depends on the value of $\rho$. Therefore, in this work 
we take three different values of $\rho$=0.75, 1.15 and 1.4. We may consider 
the effect of electric and color charge neutrality on the chiral and diquark 
condensates for the following four separate cases, of which the last two are 
new : (I) no charge neutrality condition, which was already discussed in\cite{Mandal:2009uk}; 
(II) only the color charge neutrality condition, already discussed in\cite{Fayazbakhsh:2010gc}; 
(III) only the electric charge neutrality condition is imposed; 
(IV) both the neutrality conditions are added.
\begin{figure}
\begin{center}
\includegraphics[scale=0.64]{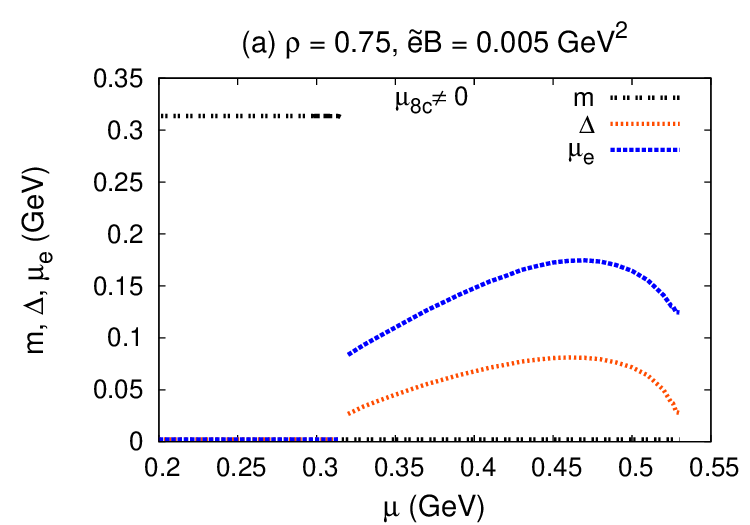}
\includegraphics[scale=0.64]{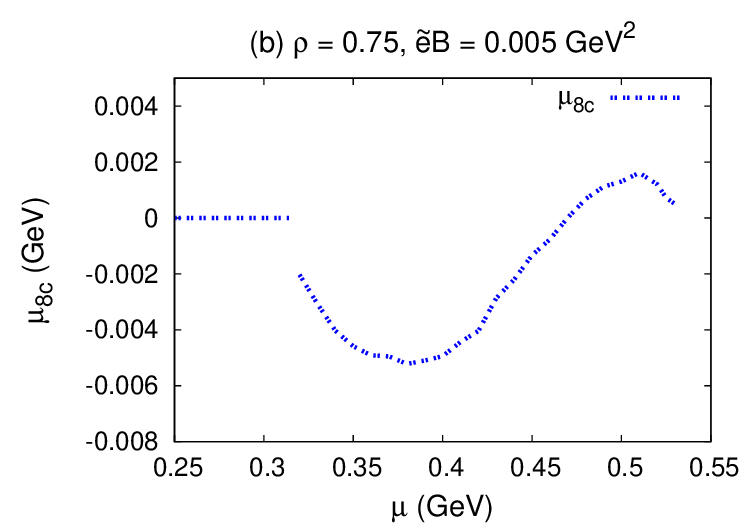}
\caption{\label{comhuang} (color online). For electric and color charge neutral 
quark matter with small magnetic field: (a) The dependence of $m$, $\Delta$ and $\mu_e$ on $\mu$
for $\tilde{e}B=0.005$ GeV$^2$ taking $\rho=0.75$ and (b) the corresponding 
dependence of $\mu_{8c}$ on $\mu$. Here we use a step function cutoff.}
\end{center}
\end{figure}

\vskip 0.2cm

Fig. \ref{comhuang}(a) shows $m,\Delta$ and the electron chemical potential 
$\mu_e$ in neutral two-flavor quark matter for $\rho=0.75$ in the weak magnetic 
field limit ($\tilde{e}B = 0.005$ GeV$^2$) - this is the g2SC phase. 
Fig. \ref{comhuang}(b) shows the small color chemical potential $\mu_{8c}$. 
We use a sharp cutoff for the results in Fig. \ref{comhuang} in order to 
compare with the zero-field results discussed in \cite{Huang:2002zd}. For 
a magnetic field $\tilde{e}B = 0.005$ GeV$^2$, $n_{max}$ is of the order of 
50 and the discrete summation becomes almost continuous. We find that our 
results for $m$, $\Delta$, $\mu_e$ and $\mu_{8c}$ in the case of $\tilde{e}B = 0.005$ GeV$^2$ are in very good agreement with the zero magnetic field results. In Fig. \ref{comhuang}(a) we see that for $\mu < \mu_c$ ($\approx 0.315$ GeV) the quark possesses a large constituent 
mass $m$ (chiral symmetry broken) and in the $\mu > \mu_c$ region, chiral 
symmetry is restored. This phenomenon is first order in nature. With $\mu_c$ 
as the critical chemical potential for this phase transition, the 
$\mu < \mu_c$ region is the chiral symmetry broken (CSB) phase. In the 
$\mu > \mu_c$ region, the diquark condensate appears, and this is
the color superconducting (CSC) phase. We find here that $\mu_c\sim m(\mu < \mu_c)$ 
and we have checked that this is true for other sets of parameters as well. 
In the rest of our analysis wherever it is not mentioned we use 
the smooth cutoff from Eq. \ref{cutoff} with $\alpha=0.01\Lambda$.

\vskip 0.2cm

\begin{figure}
\begin{center}
\includegraphics[scale=0.64]{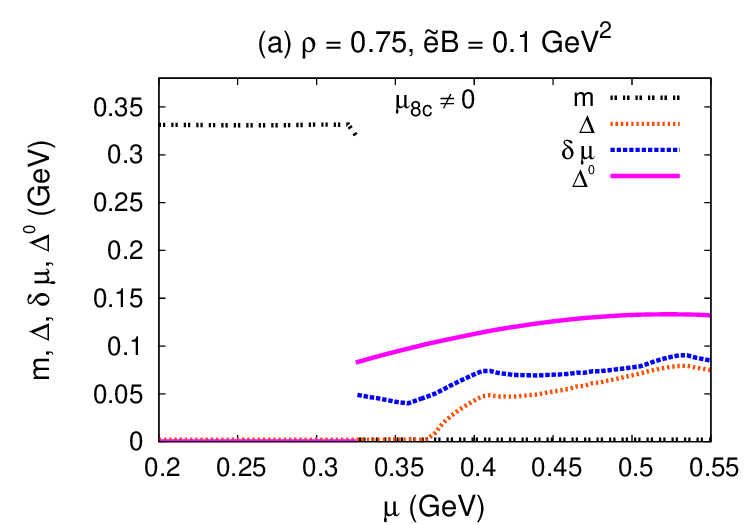}
\includegraphics[scale=0.64]{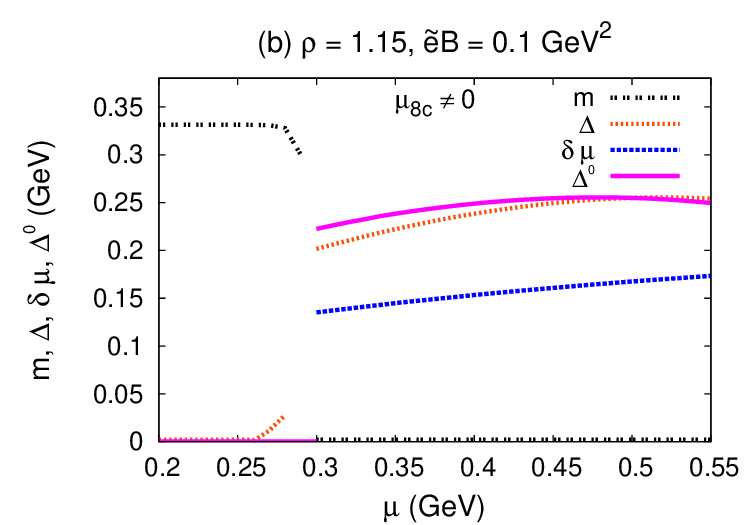}
\caption{\label{largeB} (color online). For electric and color charge neutral 
quark matter with large magnetic field: (a) The dependence of $m$, $\Delta$, 
$\delta\mu$ and $\Delta^0$ on $\mu$ for $\tilde{e}B=0.1$ GeV$^2$ taking $\rho=0.75$ 
(\emph{left panel}) and (b) $\rho=1.15$ (\emph{right panel}).}
\end{center}
\end{figure}

In Fig.~\ref{largeB} we show several physical quantities in neutral quark 
matter for large magnetic field $\tilde{e}B=0.1$ GeV$^2$, for two different 
values of $\rho$. Figs. \ref{largeB}(a) and \ref{largeB}(b) show the behavior 
of $m$, $\Delta$, $\delta\mu$ and $\Delta^0$ as a function of $\mu$ for 
$\rho=0.75$ and $\rho=1.15$ respectively. Here, $\Delta^0$ is the diquark gap 
when no charge neutrality condition is considered. In Fig. 
\ref{largeB}(a) $\delta\mu > \Delta$ and thus, this is the g2SC phase. 
The value of $\Delta$ in the neutral g2SC phase is much smaller than the 
$\Delta^0$. In Fig. \ref{comhuang}(a), the chiral and CSC phase 
transition appear to occur at the same point $\mu_c$ and both transitions are first 
order in nature. This is based on a comparison of relevant free energies~\footnote{We have checked that the thermodynamic potential with $B\neq 0$ for the neutral superconductor is indeed lower than for normal quark matter: $\Omega(\delta,\mu_e,\mu_8c) < \Omega(\delta=0,\mu_e,\mu_8c)$.}. A detailed comparison of free energies on a case-by-case basis to determine the phase diagram is ongoing and will be reported in a subsequent paper. In Fig. \ref{largeB}(a), there is a small window between 
the chiral and CSC phase transition. Though the chiral phase transition is still 
first order in nature, CSC phase transition to the g2SC phase appears to become a crossover in presence of magnetic field. We see in Fig. \ref{largeB}(b) that
$\delta\mu < \Delta$ and thus, this is expected to be the 2SC phase. In contrast to 
Fig. \ref{largeB}(a), $\Delta$ in the neutral 2SC phase is not that much
different from $\Delta^0$. 
%The chiral and superconducting phase transitions occur at the same point.
\begin{figure}
\begin{center}
\includegraphics[scale=0.64]{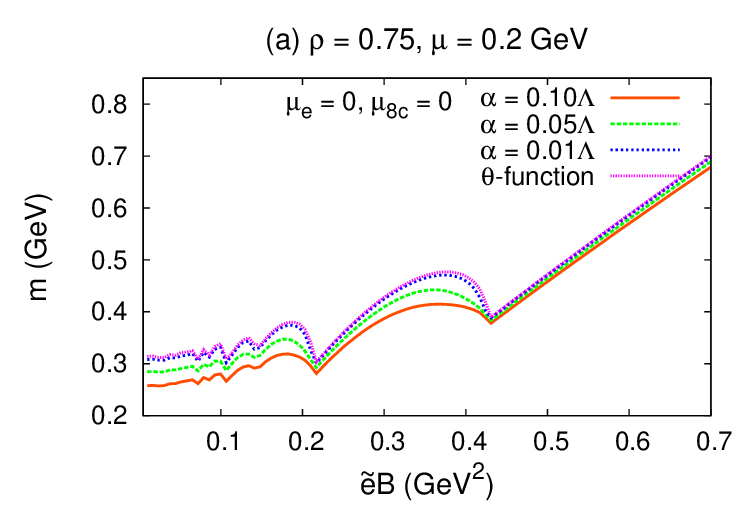}
\includegraphics[scale=0.64]{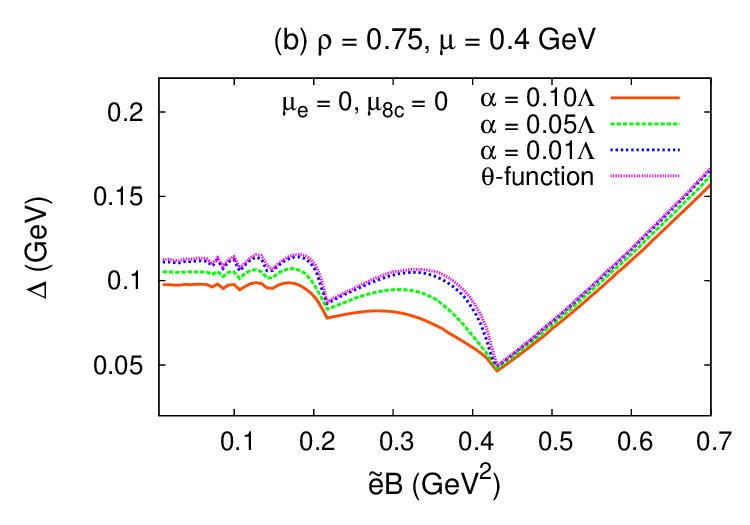}
\caption{\label{comcut} (color online). The effect of sharp/smooth cutoff 
on the chiral and the diquark gap: (a) The dependence of $m$ on $\tilde{e}B$ at $\mu=0.2$ GeV
in the chirally broken phase and (b) The dependence of $\Delta$ on $\tilde{e}B$ at $\mu=0.4$ GeV
in the superconducting phase. In both cases, neutrality conditions are not imposed. A smooth cutoff given in Eq. (\ref{cutoff}) 
with $\alpha = 0.1\Lambda$, $0.05\Lambda$, $0.01\Lambda$ and a step function cutoff are used.}
\end{center}
\end{figure}
\vskip 0.2cm
In Fig. \ref{comcut} we show the effects of choosing different cutoff schemes on $m$ and $\Delta$. In Fig. \ref{comcut}(a), we show the effect of cutoff on $m$ as a function of $\tilde{e}B$ in the CSB phase at $\mu=0.2$ GeV taking $\rho=0.75$ with no neutrality condition is imposed (case I). In 
Fig. \ref{comcut}(b), we show a similar plot for $\Delta$ in the CSC region
taking $\mu=0.4$ GeV. We choose $\mu=0.2$ GeV from the region $\mu < \mu_c$ (CSB phase) and $\mu=0.4$ GeV from the region $\mu > \mu_c$ (CSC phase) as the representative quark chemical
\begin{figure}
\begin{center}
\includegraphics[scale=0.64]{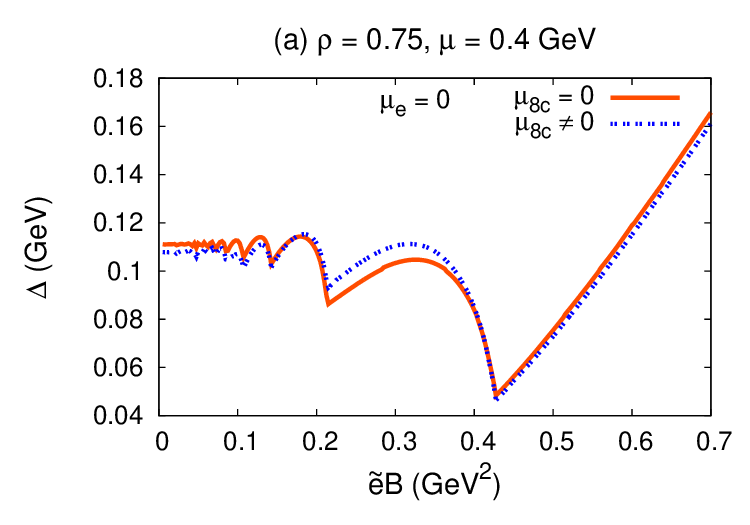}
\includegraphics[scale=0.64]{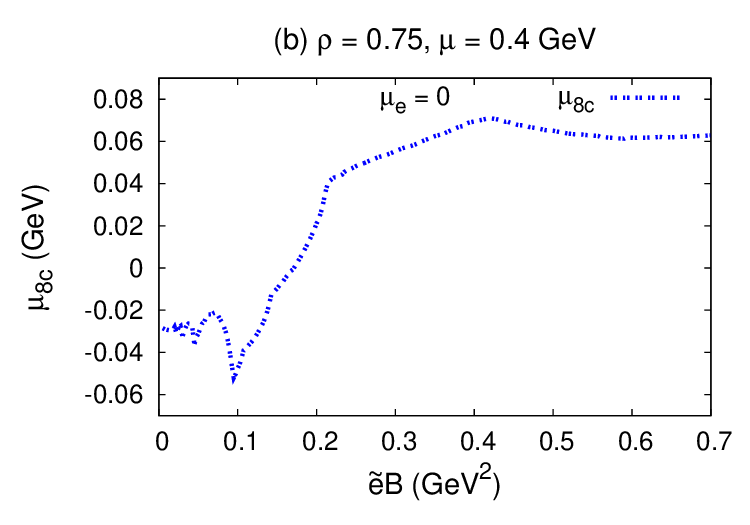}
\caption{\label{smallmuc} (color online). (a) The dependence of $\Delta$ on $\tilde{e}B$ 
at $\mu=0.4$ GeV in the CSC phase for case I and case II (see text) taking $\rho=0.75$ and 
(b) the dependence of corresponding $\mu_{8c}$ on $\tilde{e}B$ for the case II.}
\end{center}
\end{figure}
potential in these phases. For these plots we use the smooth cutoff function 
shown in Eq. (\ref{cutoff}) for three different values of sharpness parameter 
$\alpha$ and also a sharp cutoff. In all cases, we observe an oscillatory behavior 
in $m$ and $\Delta$ as a function of  $\tilde{e}B$. The ripples in the 
gap parameters become steeper for smaller values of $\alpha$. For the step 
function the ripples reach its maximum height. We find a dip in the curves 
when $\Lambda^2/\left(2|a|\tilde{e}B\right)$ takes an integer value. This kind 
of oscillatory behavior is analogous to the well known Shubnikov de Haas-van Alphen 
oscillation observed in metals in presence of magnetic field at very low 
temperature. Due to the Landau level quantization in presence of magnetic 
field the density of states and hence, various thermodynamic quantities oscillate 
as a function of magnetic field. The wavy behaviour of $m$ and $\Delta$ 
diminishes as we go to smaller values of $\tilde{e}B$. Finally $m$ and 
$\Delta$ tend toward their zero field values for sufficiently small $\tilde{e}B$ 
which corresponds to the quasi-continuous limit. We have checked for different values 
of $\rho$ and obtained similar oscillatory behavior for the gaps. Since only $G_S$ enters in the chiral gap equation, $m$ remains same for different values of $\rho$. On the other hand, $\Delta$ increases (but keeps the same oscillatory 
features) as we increase $\rho$  (or equivalently the diquark coupling strength $G_D$) since $G_D$ is directly involved in the diquark gap equation.

\begin{figure}
\begin{center}
\includegraphics[scale=0.61]{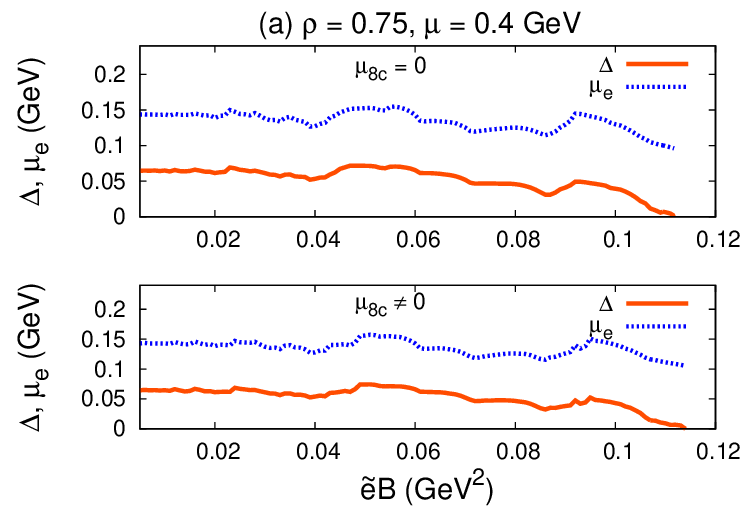}
\includegraphics[scale=0.64]{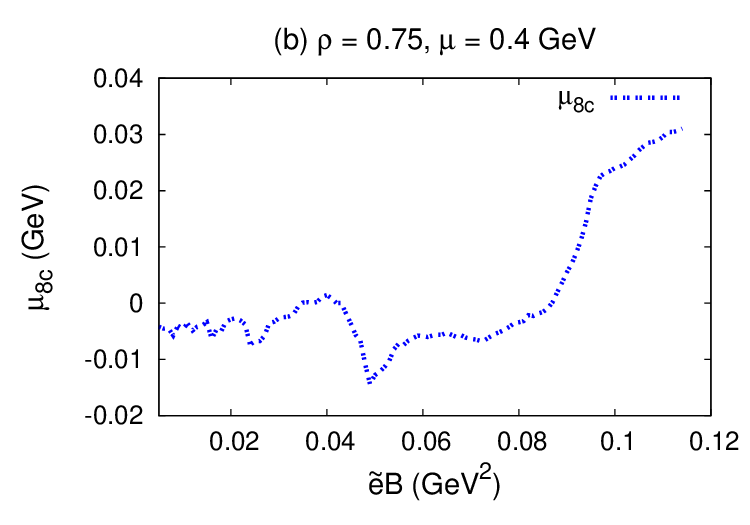}
\caption{\label{cnmag} (color online). (a) (\emph{top} and \emph{bottom panels}) The dependence of $\Delta$ and $\mu_e$ on
$\tilde{e}B$ at $\mu=0.4$ GeV in the CSC phase for the case III and case IV (see text) taking 
$\rho=0.75$ and
(b) the dependence of corresponding $\mu_{8c}$ on
$\tilde{e}B$ for case IV.}
\end{center}
\end{figure}
\begin{figure}
\begin{center}
\includegraphics[scale=0.64]{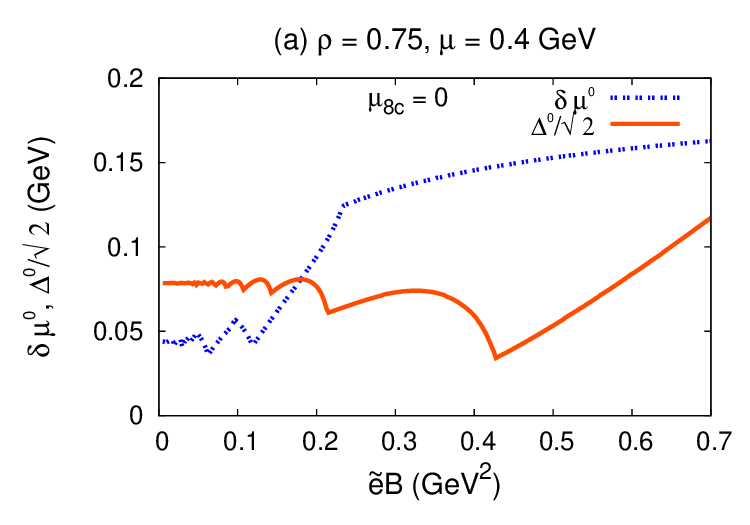}
\includegraphics[scale=0.61]{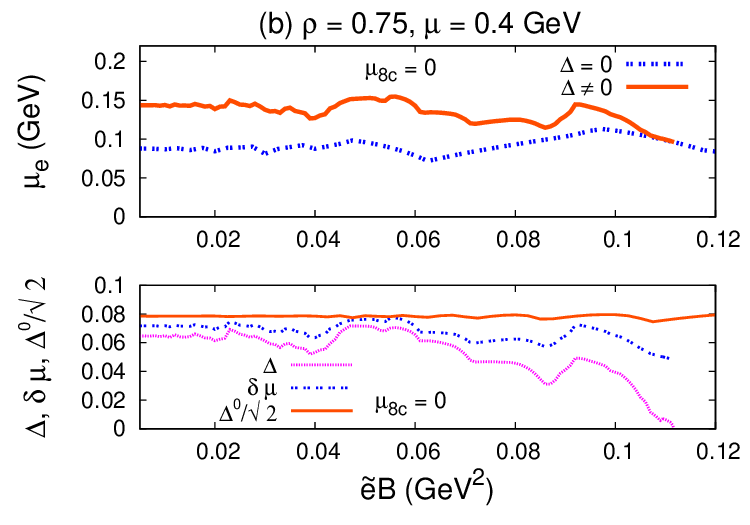}
\caption{\label{reason} (color online). (a) The dependence of $\delta\mu^0$ ($=\mu_e^0/2$) and 
$\Delta^0/\sqrt{2}$ on
$\tilde{e}B$ at $\mu=0.4$ GeV (b) (\emph{top panel}) a comparison between $\mu_e(=\mu_e^0$)
in the normal quark matter and $\mu_e$ in the CSC phase as a function of
$\tilde{e}B$ at $\mu=0.4$ GeV without $\mu_{8c}$ (b) (\emph{bottom panel}) a 
comparison between $\Delta$, $\delta\mu$ and $\Delta^0/\sqrt{2}$
as a function of $\tilde{e}B$ at $\mu=0.4$ GeV without $\mu_{8c}$. We take $\rho=0.75$ for both the plots.}
\end{center}
\end{figure}

\vskip 0.2cm

In Fig. \ref{smallmuc}(a) we show $\Delta$ as a function $\tilde{e}B$ 
in the CSC phase for case I and case II at $\mu=0.4$ GeV taking $\rho=0.75$.
We see that the effect of color neutrality ($\mu_{8c}$) on the CSC phase is 
very small. Thus, we can sometimes neglect $\mu_{8c}$ for computational 
simplicity without changing the conclusions qualitatively.
In Fig. \ref{smallmuc}(b), we show $\mu_{8c}$ as a function of
 $\tilde{e}B$ corresponding to case II of Fig. \ref{smallmuc}(a). 
In \cite{Huang:2002zd} it was shown that $\mu_{8c}$ can be positive or 
negative as a function of $\mu$. Here we find similar behavior in $\mu_{8c}$ 
as a function of $\tilde{e}B$ for a fixed value of $\mu=0.4$ GeV. In the weak 
field limit, the value of $\mu_{8c}$ is only about a few 
MeV, but its value can be fairly large 
for large values of $\tilde{e}B$. Physically, the magnetic field
creats a stress between the Fermi surfaces of the blue and the red/green
quarks.

\begin{figure}
\begin{center}
\includegraphics[scale=0.64]{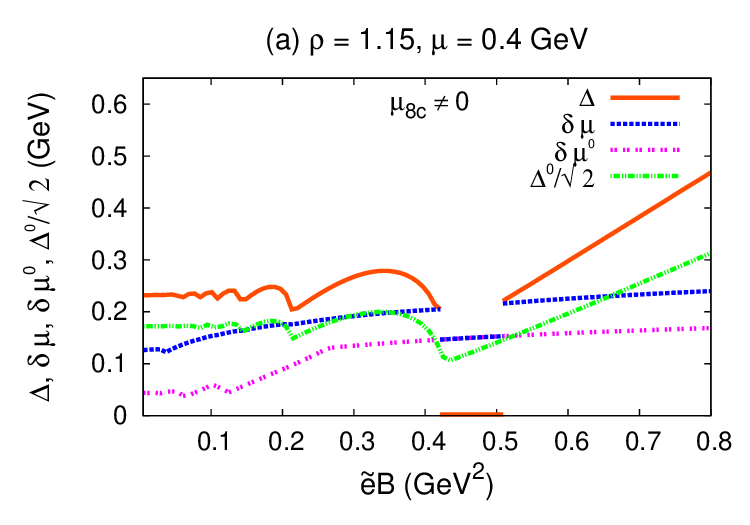}
\includegraphics[scale=0.64]{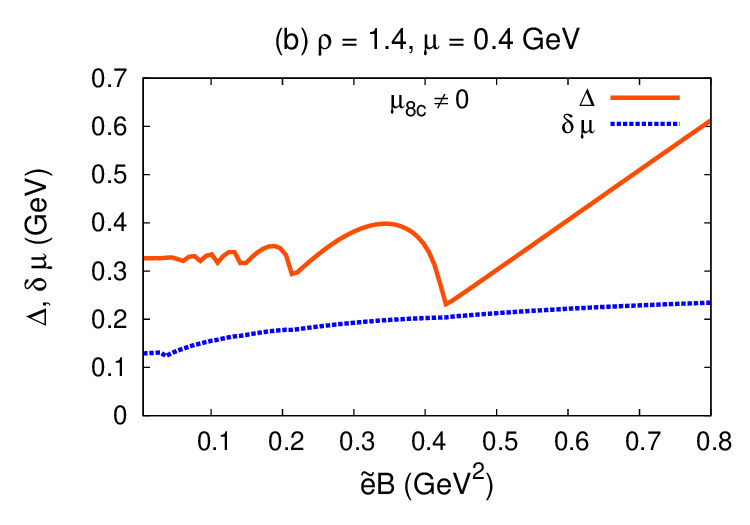}
\caption{\label{g2SC2SC} (color online). (a) The dependence of $\Delta$, $\delta\mu$, 
$\delta\mu^0$ and $\Delta^0/\sqrt{2}$ on $\tilde{e}B$ at $\mu=0.4$ GeV for case IV (see text) taking $\rho=1.15$.
(b) The dependence of $\Delta$ and $\delta\mu$ on $\tilde{e}B$ at $\mu=0.4$ 
GeV for case IV taking $\rho=1.4$.}
\end{center}
\end{figure}

\vskip 0.2cm

In Fig. \ref{cnmag}(a) we show the behavior of $\Delta$ and $\mu_e$ in 
the g2SC phase ($\rho$=0.75) as a function of $\tilde{e}B$ taking $\mu=0.4$ 
GeV. We compare the results without and with color charge neutrality 
condition imposed. Fig. \ref{cnmag}(b) shows the corresponding $\mu_{8c}$ for 
the bottom panel of Fig. \ref{cnmag}(a). The effect of $\mu_{8c}$ on $\Delta$ 
and $\mu_e$ is small, as before. But if we compare the diquark condensate 
$\Delta$ in Fig. \ref{smallmuc}(a) with that in Fig. \ref{cnmag}(a), we see 
that $\Delta$ in the charged CSC phase has a much larger value than the 
neutral case. This is because in the g2SC phase the presence of $\mu_e$ 
creates difference in the Fermi surfaces of the two pairing quarks and 
reduces the value of the diquark gap. Also in contrast to Fig. 
\ref{smallmuc}(a), enforcing charge neutrality drives $\Delta$ to zero 
at a certain value of $\tilde{e}B$. It was argued in \cite{Alford:2000ze} 
that BCS pairing is not possible if $\delta\mu^0 > \Delta^0/\sqrt{2}$. This is clearly confirmed from Fig. \ref{reason}(a) for the g2SC phase, 
where we see that $\delta\mu^0(=\mu_e^0/2$ where $\mu_e^0$ is the electron 
chemical potential in the normal quark matter) is always less than 
$\Delta^0/\sqrt{2}$ until $\Delta$ (the gap in the neutral phase) disappears, at which point $\delta\mu^0$ 
beats $\Delta^0/\sqrt{2}$. From Fig. \ref{reason}(b) (top panel), 
we also see that $\mu_e$ is larger in the CSC phase than that in neutral 
normal quark matter. At around $\tilde{e}B\sim 0.11$ GeV$^2$, $\Delta$ becomes
zero and $\mu_e$ in the CSC phase coincides with the $\mu_e$ in the neutral 
normal quark matter. Thus, the magnetic field is playing an important role in 
delineating the criterion for breakdown of the homogeneously paired phase. At this critical magnetic field, an alternate phase, such as the LOFF phase would be preferred. This is similar to the physics of the Clogston-Chandrasekhar limit predicted for electronic superconductors. An important difference is that in condensed matter systems, there is no analog of the charge neutrality condition, so $\Delta^0$ would go to zero. In case of quark matter, where QCD provides the pairing force, it is the charge neutral gap and not $\Delta^0$ that vanishes.

\vskip 0.2cm

Finally, we show in Fig. \ref{g2SC2SC} the behavior of $\Delta$ and $\delta\mu$ 
at $\mu=0.4$ GeV for two large values of $\rho=1.15$ (left panel) and 1.4 (right
panel). In presence of a magnetic field, we observe that even if $\rho>0.8$, 
quark matter is not necessarily in the 2SC phase. In Fig.~\ref{g2SC2SC}(a)
we see that most of the time $\Delta$ is larger than $\delta\mu$ which is the 
2SC phase but in the range of $\tilde{e}B\sim 0.4-0.5$ GeV$^2$, $\Delta$ 
becomes zero. The critical magnetic field value is about 4 times larger than for the g2SC phase. This is because the mismatch $\delta\mu$ in the charged CSC phase does not get as severely stressed by the magnetic field. We also note that due to the vanishing
gap, this phase can no longer be classified as the 2SC phase. In Fig.
\ref{g2SC2SC}(b) we see that for very large values $\rho=1.4$, $\Delta$ is always above $\delta\mu$ in the entire range of $\tilde{e}B$. Thus, this phase is the usual 2SC phase.

%In the region of $\tilde{e}B$ where we observe the peculiar behavior of $\Delta$, we could not find any solution for $\Delta$ except for the trivial zero solution. The corresponding $\delta\mu$ drops down to $\delta\mu^0(=\mu_e^0/2)$ where $\mu_e^0$ is the electron chemical potential in the normal quark matter phase. As we keep increasing the value of $\rho$,  the whole profile of $\Delta$ moves upwards and at some particular value of $\rho$ the whole $\Delta$-profile goes above the $\delta\mu$. 

%%%%%%%%%%%%%%%%%%%%%%%%%%%%%%%%%%%%%%%%%%%%%%%%%%%%%%%%%%%%%%%%%%%%%%%%%%%%%%%%%%

\section{Conclusions}
\label{conc}

We have studied the effect of a large magnetic field and neutrality constraints 
on the chiral and diquark condensate in a two-flavor superconductor using the NJL 
model. We used a self-consistent scheme to determine the condensates, numerically 
iterating the coupled (integral) equations for the chiral and superconducting 
gap under the constraints of color and charge neutrality. We presented results 
for the 2SC as well as the gapless 2SC phase by choosing various values of the 
$G_D/G_S$ ratio. For magnetic fields $B\lesssim 10^{18}$G, our results for the 
chiral and superconducting gaps align with the zero magnetic field results of
~\cite{Huang:2002zd}, since many Landau levels ($\approx 50$) are being summed 
over, which approximates the continuum result. We note that the chiral and 
superconducting phase transitions occur at the same value of $\mu_q$.
This changes at very large values of $B\geq 10^{18}$G for the gapless 2SC phase, 
where the color superconducting phase transition becomes a crossover. Furthermore, 
the value of the superconducting gap in the gapless 2SC phase is much smaller than 
that in the charge neutral 2SC phase. Both effects have to do with stresses on the 
up and down quark Fermi surfaces due to the charge neutrality constraint.

\vskip 0.2cm

As found in previous works on the three-flavor gap or CFL phase~\cite{Noronha:2007wg,Noronha:2007ps}, 
the two-flavor gap at large magnetic field $B\geq 10^{18}$G exhibits Shubnikov de Haas-van Alphen 
oscillations due to discrete fluctuations of the quark density of states as a function of 
magnetic field. These oscillations can become quite large for strong magnetic 
fields, and even for $G_D/G_S$ ratios that normally correspond to strong coupling, 
i.e, yield the  2SC phase at zero field, the gap can vanish when $\delta\mu>\Delta$. 
Physically, this means that the homogeneous gap ansatz is not the correct one anymore, 
and probably a LOFF phase~\cite{Ful,LO} is the preferred one. Large oscillations in the 
gap can also lead to the formation of magnetic domains if the homogeneous phase becomes 
unstable/metastable, with applications to magnetar phenomena as argued in~\cite{Noronha:2007wg}. 
Another interesting consequence of charge neutrality, through the momentum mismatch 
of the down and up quarks, is the Clogston-Chandrasekhar limit~\cite{Clogs,Chand}, 
originally formulated as a maximum critical magnetic field in hard electronic superconductors. 
While difficult to confirm experimentally in metals, due to the strong Meissner effect, 
it can be achieved more easily in a color superconductor, where the rotated photon 
$\tilde{\gamma}$ can penetrate. In the gapless 2SC phase, we find indications that this limit 
is reached at $B\approx 2\times 10^{19}$G, but the magnetic field imposes oscillations in the gap that preclude a definitive conclusion about the order of the transition. It should be noted, however, that it is 
not the magnetic field that destroys the BCS state. In fact, without electric charge 
neutrality enforced, the magnetic field would actually increase the value of the gap, 
since the magnetic moments of the pairing quarks are aligned. Rather, the mismatch 
of Fermi surfaces increases with increasing magnetic field, forcing the gap in the (locally) charge neutral phase to get 
smaller. Thus, the magnetic field has an essential role to play in delineating the breakdown of the homogeneous pairing ansatz. 

\vskip 0.2cm

Finally, although we have used an effective model with a cutoff, which causes some uncertainty in the calculated values of the quark gap and constituent mass, it is interesting to speculate on possible physical consequences of the large magnetic field, and its effect on the condensates. At some fixed large value of the local field $B$ and in the density window of the metastable regions mentioned above, cooling of two-flavor quark matter below the superconducting transition temperature $T_c$ can result in the formation of domains or nuggets of superconducting regions with different values for the gap, even at constant pressure (a detailed analysis requires assessment 
of surface and screening effects in superconducting quark matter~\cite{Jaikumar:2005ne,Alford:2006bx}). 
We note that once neutrality is imposed, there is no range of density where the chiral and diquark gap co-exist. The mixed broken phase found in~\cite{Huang:2001yw} does not survive under physical conditions of neutrality. However, in the chirally broken phase (where the diquark gap is zero), due to the magnetic field, even the chiral gap (or constituent mass) oscillates. We can also imagine that external magnetic fields of order $eB/\mu^2\sim 1$ show some local variation on the microscopic scale in the initial stages of the formation of the dense neutron star. In this case, magnetic domains with different magnetization can form. At lower density, domains of broken chiral symmetry can also form, 
hearkening back to the disordered chiral condensate (DCC) idea~\cite{Rajagopal:1993ah}. Such kinds of nucleation and domain formation will release latent heat that might be very large owing to the large value of the magnetic field, serving as a large internal engine for possible energetic events on the surface of the neutron star. Such internal mechanisms are unlikely to occur in a pure neutron star without a quark core~\cite{Broderick:2001qw} and could have important applications for astrophysics of gamma-ray bursts~\cite{Ouyed:2005dz}. 
In addition, it has been recently shown that large magnetic fields in strange quark matter introduce anisotropies in bulk viscosities, which can change the stability region of the r-mode in such stars~\cite{Huang:2009ue} and the epoch of gravitational wave emission as the star spins down. A similar effect may arise in magnetized superconducting two-flavor quark matter, pointing to another important link between the microscopic and astrophysical consequences of a large magnetic field in quark matter in the core of neutron stars.

\section*{Acknowledgments}

We are grateful to Dr. Sanatan Digal for many fruitful discussions and to Neeraj Kumar Kamal with assistance in certain numerical aspects of this work.

\end{document}